\documentclass[twocolumn]{aastex631}

\usepackage{graphicx}
\usepackage[varg]{txfonts}
\usepackage{cases}
\usepackage{natbib}
\usepackage{multirow}
\usepackage{longtable}
\usepackage{booktabs}

\shorttitle{Energetics of a flare and a CME generated by a HC eruption}
\shortauthors{Zhang et al.}

\graphicspath{{./}{figures/}}

\begin{document}

\title{Energetics of a solar flare and a coronal mass ejection generated by a hot channel eruption}

\correspondingauthor{Qingmin Zhang}
\email{zhangqm@pmo.ac.cn}

\author[0000-0003-4078-2265]{Qingmin Zhang}
\affiliation{Key Laboratory of Dark Matter and Space Astronomy, Purple Mountain Observatory, CAS, Nanjing 210023, People's Republic of China}
\affiliation{Yunnan Key Laboratory of Solar physics and Space Science, Kunming 650216, People's Republic of China}

\author[0000-0002-5625-1955]{Weilin Teng}
\affiliation{Department of Astronomy and Space Science, University of Science and Technology of China, Hefei 230026, People's Republic of China}
\affiliation{Key Laboratory of Dark Matter and Space Astronomy, Purple Mountain Observatory, CAS, Nanjing 210023, People's Republic of China}

\author[0000-0002-4538-9350]{Dong Li}
\affiliation{Key Laboratory of Dark Matter and Space Astronomy, Purple Mountain Observatory, CAS, Nanjing 210023, People's Republic of China}
\affiliation{Yunnan Key Laboratory of Solar physics and Space Science, Kunming 650216, People's Republic of China}

\author[0000-0003-4787-5026]{Jun Dai}
\affiliation{Key Laboratory of Dark Matter and Space Astronomy, Purple Mountain Observatory, CAS, Nanjing 210023, People's Republic of China}

\author[0000-0003-1979-9863]{Yanjie Zhang}
\affiliation{Key Laboratory of Dark Matter and Space Astronomy, Purple Mountain Observatory, CAS, Nanjing 210023, People's Republic of China}

\begin{abstract}
Hot channels (HCs) are prevalent in the solar corona and play a critical role in driving flares and coronal mass ejections (CMEs).
In this paper, we estimate the energy contents of an X1.4 eruptive flare with a fast CME generated by a HC eruption on 2011 September 22.
Originating from NOAA active region 11302, the HC is the most dramatic feature in 131 and 94 {\AA} images observed by
the Atmospheric Imaging Assembly (AIA) on board the Solar Dynamics Observatory (SDO).
The flare is simultaneously observed by SDO/AIA, the Reuven Ramaty High-Energy Solar Spectroscopic Imager (RHESSI),
and the Extreme-Ultraviolet Imager (EUVI) on board the behind Solar TErrestrial RElations Observatory (STEREO).
The CME is simultaneously detected by the white-light coronagraphs of the Large Angle Spectroscopic Coronagraph (LASCO) on board SOHO and the
COR1 coronagraph on board the behind STEREO. Using multiwavelength and multiview observations of the eruption, various energy components of the HC, flare, and CME are calculated.
The thermal and kinetic energies of the HC are (1.77$\pm$0.61)$\times10^{30}$ erg and (2.90$\pm$0.79)$\times10^{30}$ erg, respectively.
The peak thermal energy of the flare and total radiative loss of SXR-emitting plasma are (1.63$\pm$0.04)$\times10^{31}$ erg 
and (1.03$-$1.31)$\times10^{31}$ erg, respectively. The ratio between the thermal energies of HC and flare is 0.11$\pm$0.03, suggesting that thermal energy of the HC is not negligible.
The kinetic and potential energies of the CME are (3.43$\pm$0.94)$\times10^{31}$ erg and (2.66$\pm$0.49)$\times10^{30}$ erg, 
yielding a total energy of (3.69$\pm$0.98)$\times10^{31}$ erg for the CME.
Continuous heating of the HC is required to balance the rapid cooling by heat conduction, which probably originate from intermittent magnetic reconnection at the flare current sheet.
Our investigation may provide insight into the buildup, release, and conversion of energies in large-scale solar eruptions.
\end{abstract}


\section{Introduction} \label{intro}
A magnetic flux rope (MFR) is composed of a group of twisted magnetic field lines wrapping around a common axis \citep[see][and references therein]{cx17,liu20}. 
The helical structure of a MFR naturally provides dips to support the dense materials in solar prominences \citep{vb89,oka08,zyh18,jk21,lia23}. 
MFRs play a central role in driving large-scale solar eruptions, leading to flares and coronal mass ejections \citep[CMEs;][]{chen11}.
MFRs are involved in many theoretical models of eruptions, such as the ideal kink instability \citep{tor05}, flux emergence model \citep{chen00}, catastrophic loss of equilibrium model \citep{lf00},
tether-cutting model \citep{mo01,jia21}, and breakout model \citep{lyn04}.
On the solar disk, MFRs may appear as sigmoids with a forward-$\mathsf{S}$ shape \citep{au10,gre11,sav12,jam17} or an inverse-$\mathsf{S}$ shape \citep{rust96,liur10,sav16}.
Above the solar limb, MFRs show up in extreme ultraviolet (EUV) and white light (WL) wavelengths \citep{dere99,yan18,zqm22a}.
Occasionally, multiple MFRs are found to coexist in an active region (AR) and erupt sequentially \citep{awa18,hou18,wy18}. 
MFRs are mainly formed in the corona via magnetic reconnection between two $\mathsf{J}$-shaped sheared arcades before or during eruptions \citep{vb89,ama03,xia14,xue17}. 
Interaction between two sheared arcades is characterized by transient brightening \citep{jos15} and bidirectional outflows \citep{cx15,yan21}.
The line-tied footpoints of MFRs may experience drifting motion due to continuous magnetic reconnection in the corona \citep{au19}.
\citet{xc20} proposed a new method and explored the evolution of toroidal flux of CME flux ropes during their eruptions.
The toroidal flux of the CME flux rope rapidly increases before decreasing in a gradual way, which is similar to the evolution of SXR flux of the associated flare.

Hot channels (HCs) are bright structures first observed in 94 {\AA} ($\log T[\mathrm{K}]\approx6.85$) and 131 {\AA} ($\log T[\mathrm{K}]\approx7.05$) 
of the Atmospheric Imaging Assembly \citep[AIA;][]{lem12} on board the Solar Dynamics Observatory (SDO) spacecraft \citep{cx11,zj12}. 
Due to the extraordinarily high temperatures, HCs are hardly visible in other EUV passbands of AIA \citep{cx12,zqm22b}.
Composite images show that HCs are surrounded by cooler leading edges or compression fronts observed in 171, 193, and 211 {\AA} during eruptions.
Apart from hot EUV and SXR passbands, HCs could be detected in microwaves \citep{wu16}.
It is widely accepted that the magnetic configurations of HCs are MFRs, and HCs play a key role in generating CMEs \citep{cx13a,nin15}.
HCs could be created before \citep{cx13a,zqm22b} or during eruptions \citep{cx11,song14,chen19,gou19,liu22}.
Statistical investigation reveals that $\sim$2/3 and $\sim$1/3 of them are formed before and during eruptions \citep{nin20}, respectively.

\begin{figure}
\includegraphics[width=0.45\textwidth,clip=]{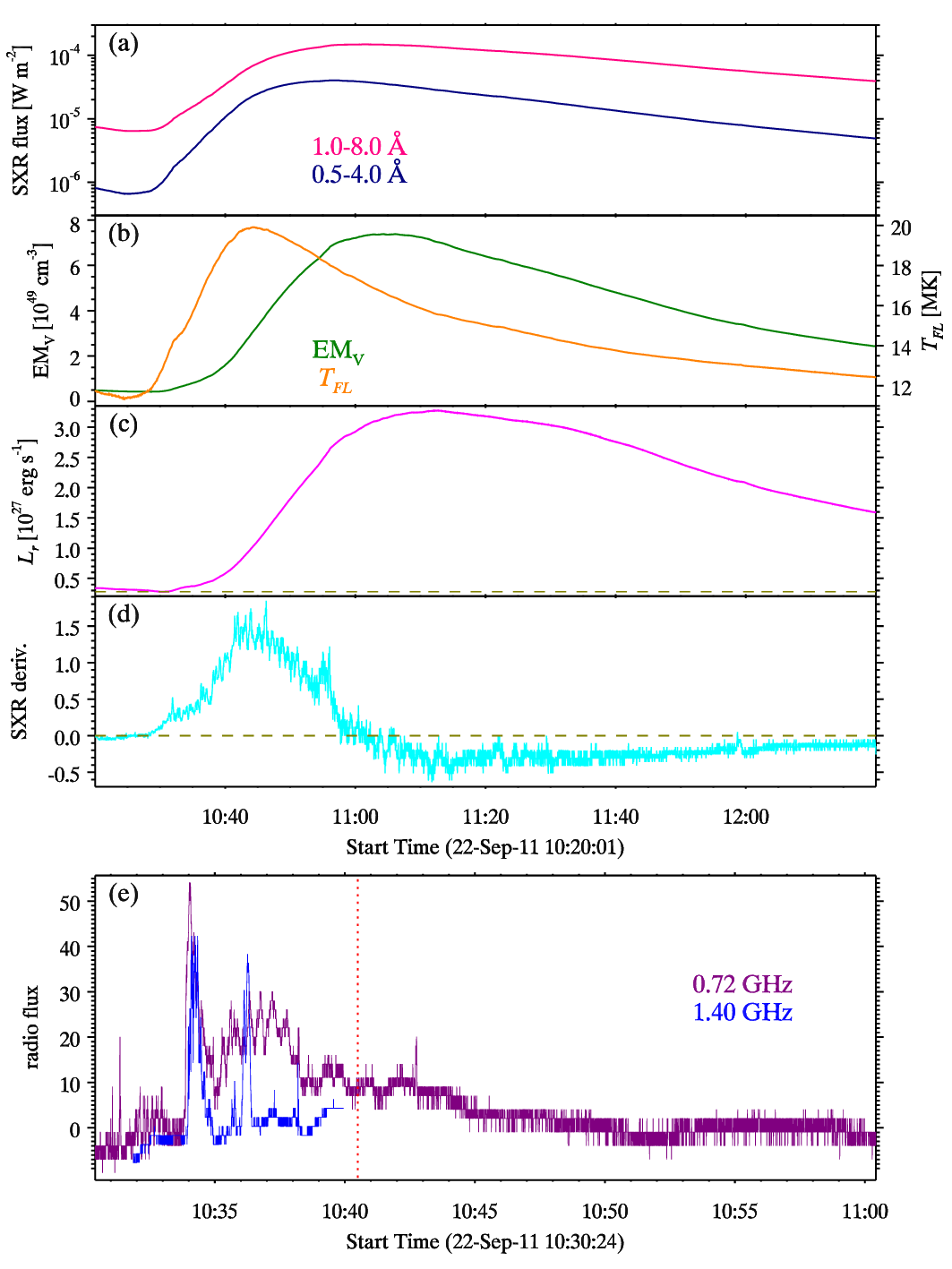}
\centering
\caption{(a)-(d) Time evolutions of SXR fluxes in 1$-$8 {\AA} (pink line) and 0.5$-$4 {\AA} (dark blue line), volume emission measure (green line) and temperature (orange line), 
radiative loss rate of SXR-emitting plasma (magenta line), and SXR flux derivative (cyan line) of the X1.4 flare, respectively.
(e) Time evolutions of the radio fluxes at 0.72 GHz (purple line) and 1.40 GHz (blue line) during the impulsive phase of the flare.
The red dotted line denotes the time at 10:40:30 UT when $v_{HC}$ reaches the linear speed of CME.}
\label{fig1}
\end{figure}

Although HCs have drawn great attention in the past decade, the energetics of HCs is rarely explored.
Using the differential emission measure (DEM) analysis \citep{hk12}, \citet{hk13} studied the erupting plasmoid on 2010 November 3 and calculated its temperature and density.
After dividing the plasmoid into a core, an envelope, and a long stem (current sheet), 
they calculated thermal energy ($\sim$2.6$\times10^{30}$ erg) and kinetic energy ($\sim$1.6$\times10^{30}$ erg) of the plasmoid, respectively.
Hence, the total energy including the two components reaches $\sim$4.2$\times10^{30}$ erg. 
Note that the plasmoid core with temperatures of 11$-$14 MK is considered as the edge-on representative of the HC \citep{cx11}.
Since the associated flare was occulted by the eastern limb, they did not calculate the flare energy.
In this paper, using multiwavelength and multiview observations on 2011 September 22, we study the eruption of a HC, which generates an X1.4 class flare and a fast halo CME in AR 11302 (N11E80).
The flare commences at $\sim$10:29 UT, peaks at $\sim$11:04 UT, and stops after 12:00 UT (Figure~\ref{fig1}(a)).
We will estimate different energy components of the flare, CME, and HC. The paper is organized as follows. 
We describe observations and data analysis in Section~\ref{data}. The energy components are obtained in Section~\ref{eng} and compared with previous results in Section~\ref{dis}.
Finally, a brief conclusion is given in Section~\ref{con}.

\section{Observations and data analysis} \label{data}
\begin{figure*}
\includegraphics[width=0.90\textwidth,clip=]{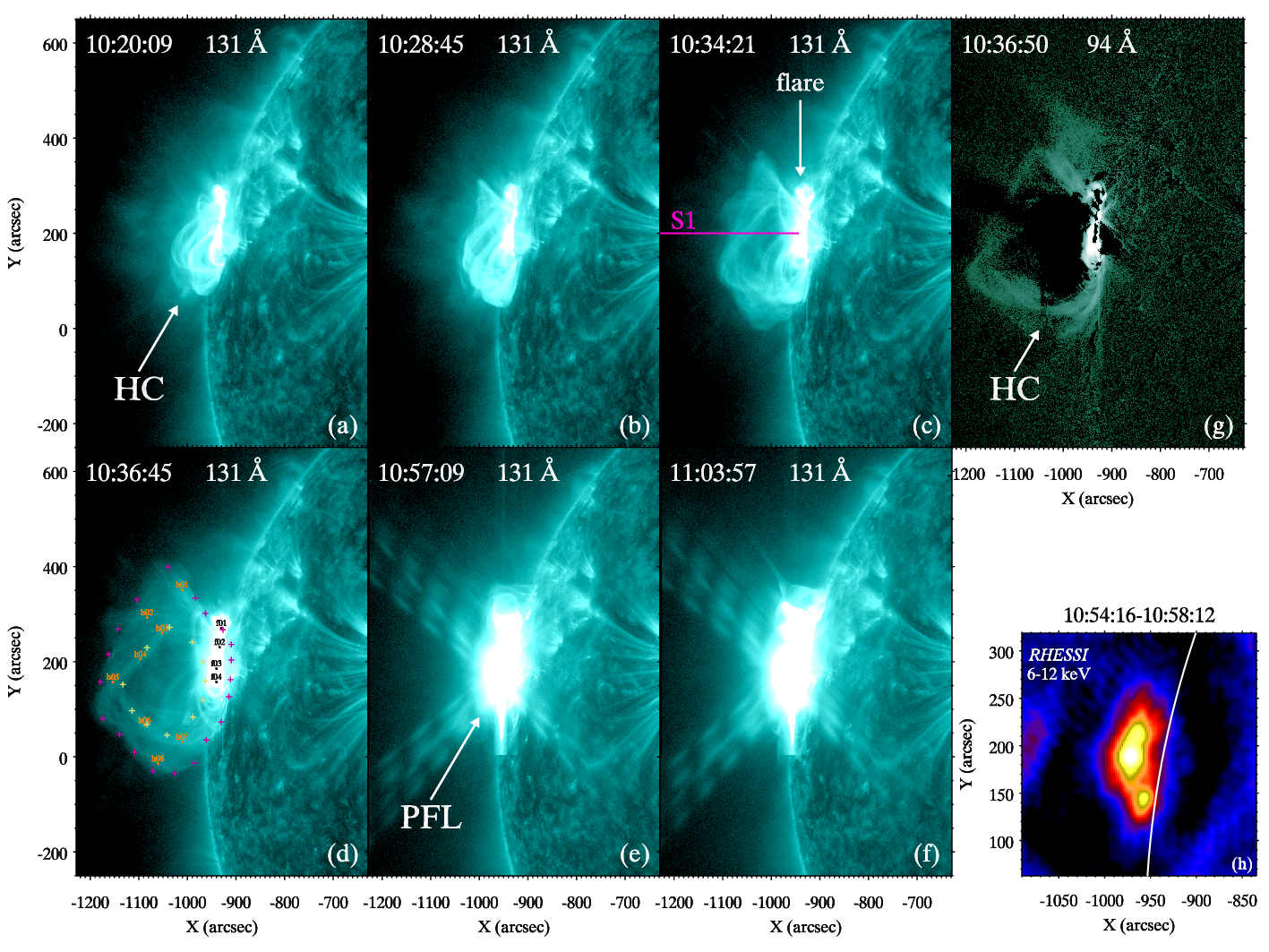}
\centering
\caption{(a)-(f): AIA 131 {\AA} images to illustrate the evolution of HC and flare. In panel (c), a horizontal slice (S1) with a length of 206.6 Mm is used to investigate the evolution of HC.
In panel (d), the magenta and yellow ``+" symbols represent the outer and inner boundaries of the HC and flare.
Eight orange boxes (h01-h08) and four black boxes (f01-f04) are used to perform DEM analysis for the HC and flare, respectively.
(g) AIA base-difference image in 94 {\AA} at 10:36:50 UT. 
(h) SXR image of the flare at 6$-$12 keV obtained with RHESSI during 10:54$-$10:58 UT.
An animation showing the HC eruption in AIA 131 and 94 {\AA} is available.
It covers a duration of 44 minutes from 10:20 UT to 11:04 UT on 2011 September 22. The entire movie runs for $\sim$3 s.
(An animation of this figure is available.)}
\label{fig2}
\end{figure*}

Figure~\ref{fig2}(a)-(f) show six AIA 131 {\AA} images during 10:20$-$11:04 UT (see also the animation). Panel (a) features the HC before eruption, which is pointed by the arrow.
The HC lifts off slowly until $\sim$10:28 UT (panel (b)). Afterwards, it starts to rise impulsively, generating the X1.4 flare below (panel (c)).
The HC continues to ascend and expand laterally (panel (d)), leaving behind hot and bright post flare loops (PFLs) close to the flare peak time (panels (e)-(f)).
The rising HC is also evident in 94 {\AA} base-difference image in Figure~\ref{fig2}(g).
Figure~\ref{fig2}(h) shows soft X-ray (SXR) image of the flare at 6$-$12 keV, which is obtained using the CLEAN method \citep{hur02} during 10:54:16$-$10:58:12 UT
with the Reuven Ramaty High-Energy Solar Spectroscopic Imager \citep[RHESSI;][]{lin02}. The integration time is 236 s and the detectors are 3F, 4F, 5F, 6F, 7F, and 8F.
Since emission at 6$-$12 keV is mainly from thermal electrons, the shape of the flare is very close to that in 131 {\AA} (panel (e)).

\begin{figure}
\includegraphics[width=0.45\textwidth,clip=]{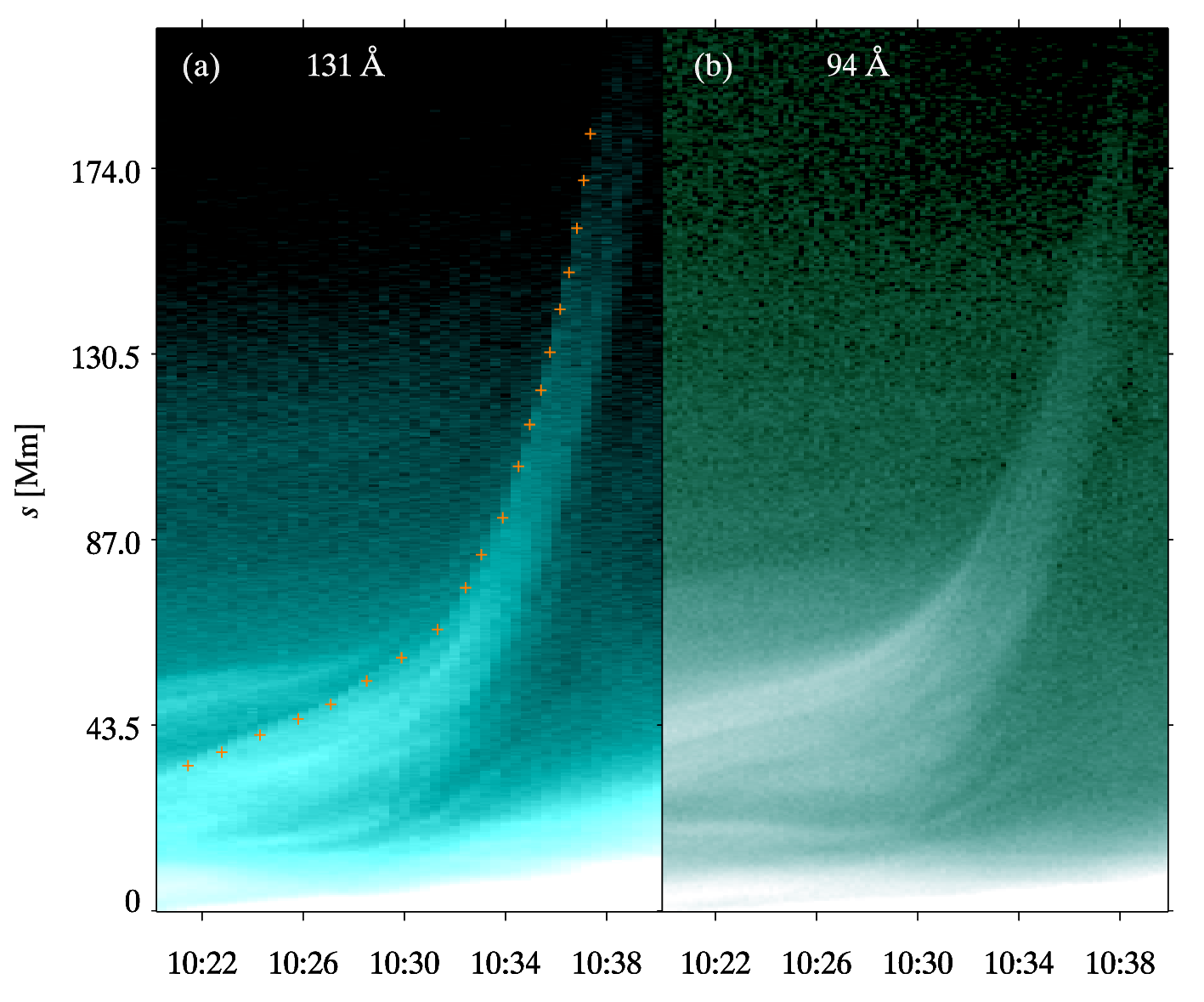}
\centering
\caption{Time-distance diagrams of S1 in 131 and 94 {\AA}. $s=0$ and $s=206.6$ Mm denote the west and east endpoints of S1, respectively.}
\label{fig3}
\end{figure}

\begin{figure}
\includegraphics[width=0.45\textwidth,clip=]{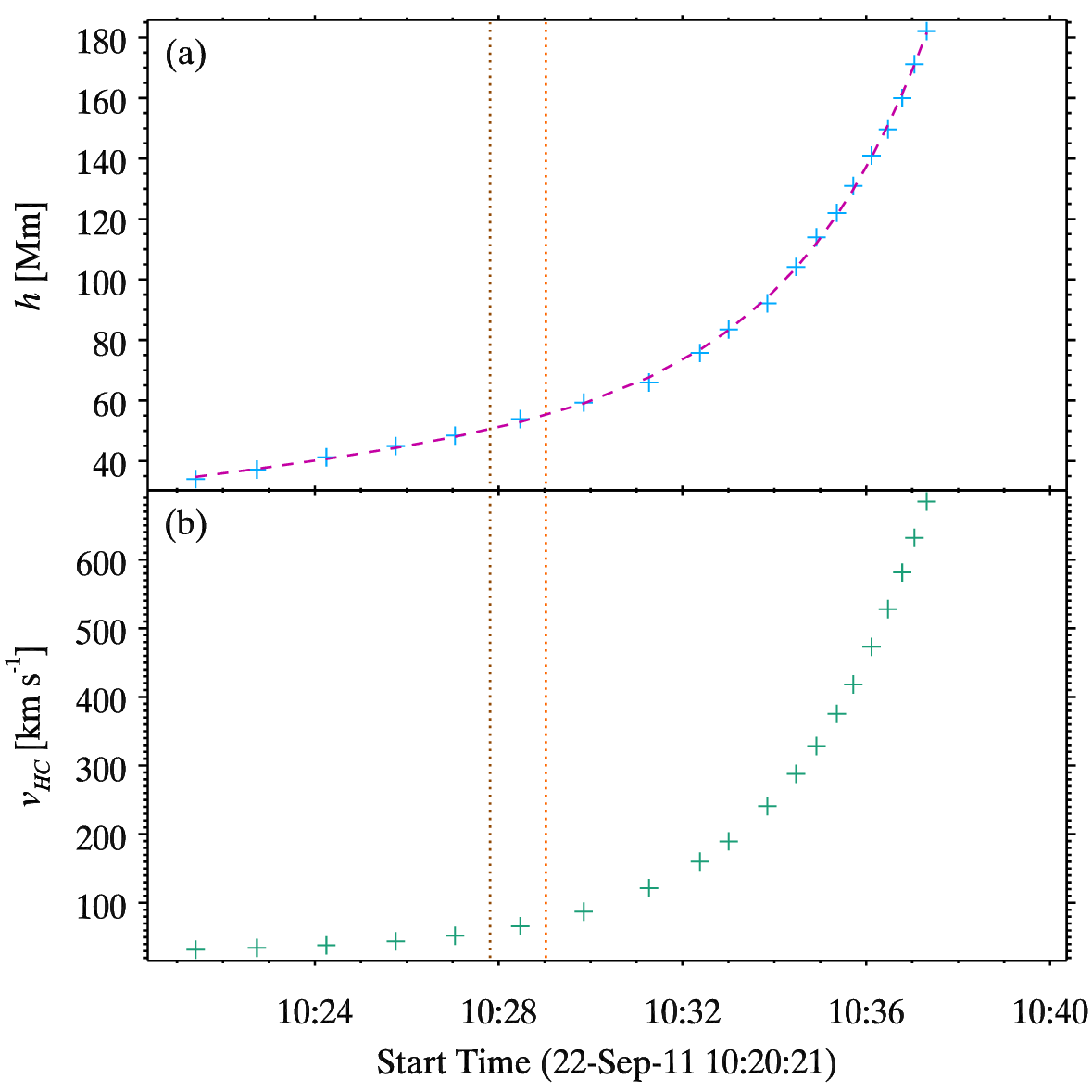}
\centering
\caption{Time evolutions of the height (a) and velocity (b) of the HC. The pink dashed line delineates the result of curve fitting using Equation~(\ref{eqn-1}).
The brown and orange dotted lines represent the onset time of fast rise and start time of the flare, respectively.}
\label{fig4}
\end{figure}

In order to track the HC, a horizontal slice (S1) is selected along the direction of eruption in Figure~\ref{fig2}(c). Time-distance diagrams of the slice in 131 and 94 {\AA} are displayed in Figure~\ref{fig3}.
The evolution of the HC is similar in both passbands. 
In panel (a), the trajectory of the HC is marked with orange ``+" symbols, which is characterized by a slow rise followed by a fast rise \citep{cx13b}.
In Figure~\ref{fig4}(a), the same trajectory is marked with blue ``+" symbols.
To fit the trajectory, we use the equation \citep{cx13b}:
\begin{equation} \label{eqn-1}
  h(t)=c_{0}e^{t/\sigma}+c_{1}t+c_{2},
\end{equation}
where $t$ denotes time after $t_{0}$ (10:20:21 UT), $h(t)$ is height of the HC, and $\sigma$, $c_{0}$, $c_{1}$, and $c_{2}$ are free parameters.
The onset time of fast rise is expressed as: 
\begin{equation} \label{eqn-2}
  t_{\mathrm{onset}}=\sigma\ln(\frac{c_{1}\sigma}{c_{0}}).
\end{equation}
 
In Figure~\ref{fig4}, the time of $t_{\mathrm{onset}}$ at 10:27:48 UT (brown dotted line) is $\sim$72 s earlier than the flare start time at 10:29:00 UT (orange dotted line),
suggesting that the HC is formed before flare \citep{nin20}.
Evolution of the associated velocity ($v_{HC}=\frac{dh(t)}{dt}$) is displayed in Figure~\ref{fig4}(b), 
showing that $v_{HC}$ increases slowly from $\sim$30 to $\sim$50 km s$^{-1}$ at $t_{\mathrm{onset}}$ and increases impulsively to $\sim$680 km s$^{-1}$ around 10:37 UT.
\citet{cx20} explored the early kinematic evolutions of 12 eruption events, including the HC eruption on 2011 September 22 (H3 in their list).
Time-distance diagram of the HC in 131 {\AA} is derived during 10:10$-$10:44 UT (see their Fig. 2(a)).
Trajectory of the HC is extracted during 10:16$-$10:38 UT, which is fitted with various functions (see their Fig. 3(a)).
The onset time of main acceleration obtained from Equation~(\ref{eqn-1}) is $\sim$10:25 UT, which is $\sim$3 minutes earlier than our result.
There are two reasons for the inconsistency. Firstly, the trajectory of the HC is derived manually rather than automatically in both works. Hence, uncertainty of the HC leading edge is inevitable.
Secondly, the start time of HC evolution is set to be 10:16:00 UT in \citet{cx20}, which is $\sim$4 minutes earlier than that in our study (see Figure~\ref{fig4}).
Nevertheless, the inconsistency does not influence the calculation of the energetics of HC.

\begin{figure}
\includegraphics[width=0.45\textwidth,clip=]{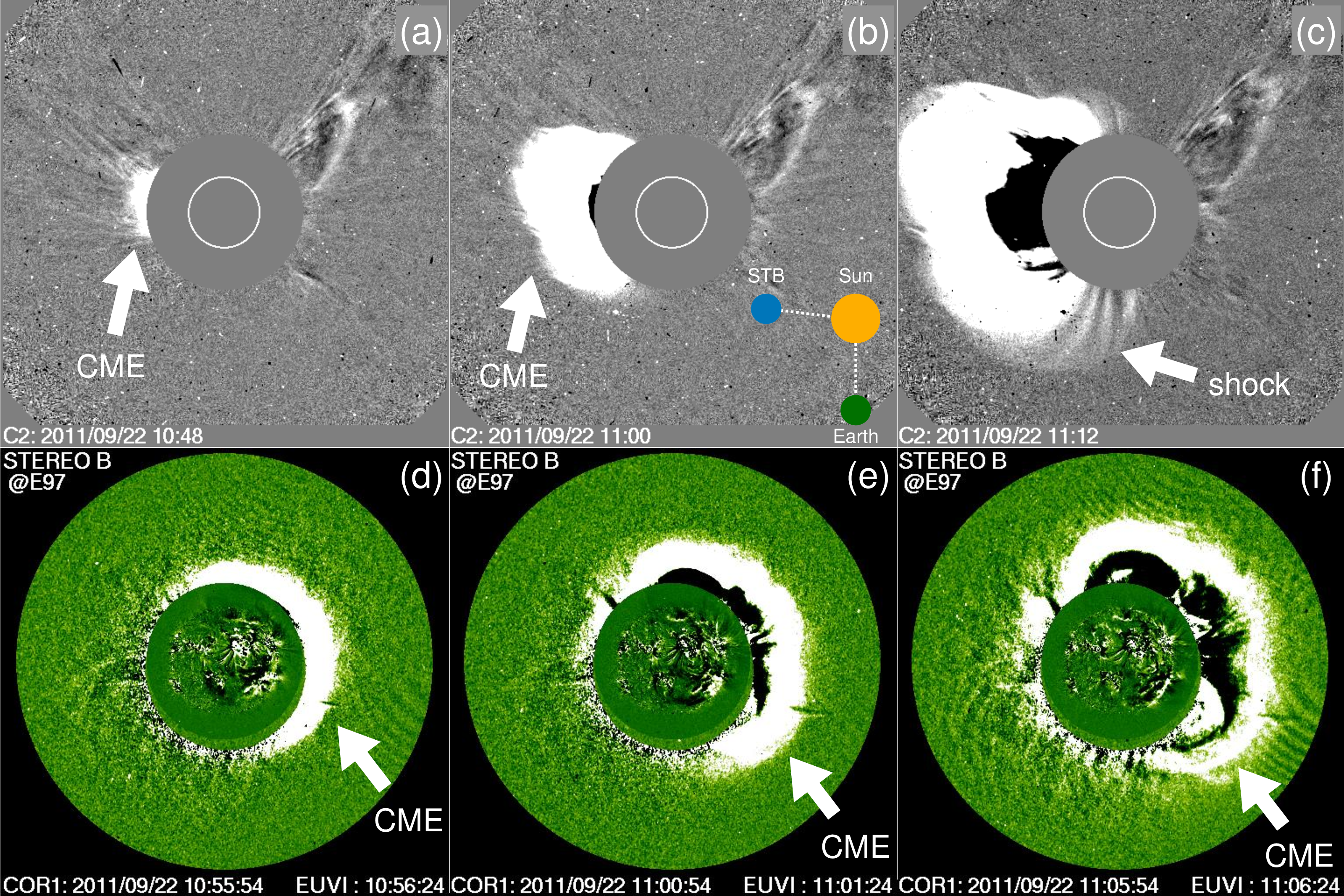}
\centering
\caption{Running-difference images of the CME observed by LASCO/C2 during 10:48$-$11:12 UT (top panels) and STB/COR1 during 10:56$-$11:06 UT (bottom panels).
The white arrows point to the CME and CME-driven shock.}
\label{fig5}
\end{figure}

In Figure~\ref{fig5}, the top and bottom panels show WL images of the related CME observed by C2 coronagraph of the Large Angle Spectroscopic Coronagraph \citep[LASCO;][]{bru95} 
on board SOHO \footnote{https://cdaw.gsfc.nasa.gov/CME\_list/} and by COR1 coronagraph on board the behind Solar TErrestrial RElations Observatory \citep[STEREO;][]{kai08}, respectively.
The behind STEREO (hereafter STB) has a separation angle of 96.5$\degr$ with respect to the Sun-Earth connection on 2011 September 22.
The CME first appears in the field of view (FOV) of LASCO/C2 at 10:48 UT and propagates eastward with a position angle of $\sim$71$\degr$.
It is noted that the CME does not present a typical three-part structure \citep{song23a,song23b,zyh23}. Instead, it features a bright leading front without a trailing core.
According to Equation~(\ref{eqn-1}), the HC rises to a height of $\sim$410 Mm above the limb at 10:40:30 UT, which is still blocked by the occulting disk of LASCO/C2.
The corresponding velocity ($v_{HC}$) increases to $\sim$1909 km s$^{-1}$, which is close to the linear speed of CME.
Assuming a constant speed after 10:40:30 UT, the HC should reach a height of $\sim$2.84 $R_{\sun}$ at 10:48:06 UT, 
which is very close to the height of CME leading edge, i.e., $\sim$2.98 $R_{\sun}$.
In light of the direction and kinematics, it is most likely that the eruptive HC evolves into the CME leading edge \citep{vour13}.
The fast CME drives a shock wave indicated by the arrow in panel (c), which is accompanied by a type II radio burst \citep{zuc14}.
In the FOV of STB/COR1, since the source region (AR 11302) is close to the disk center, the CME is full halo without a core.

\begin{figure}
\includegraphics[width=0.45\textwidth,clip=]{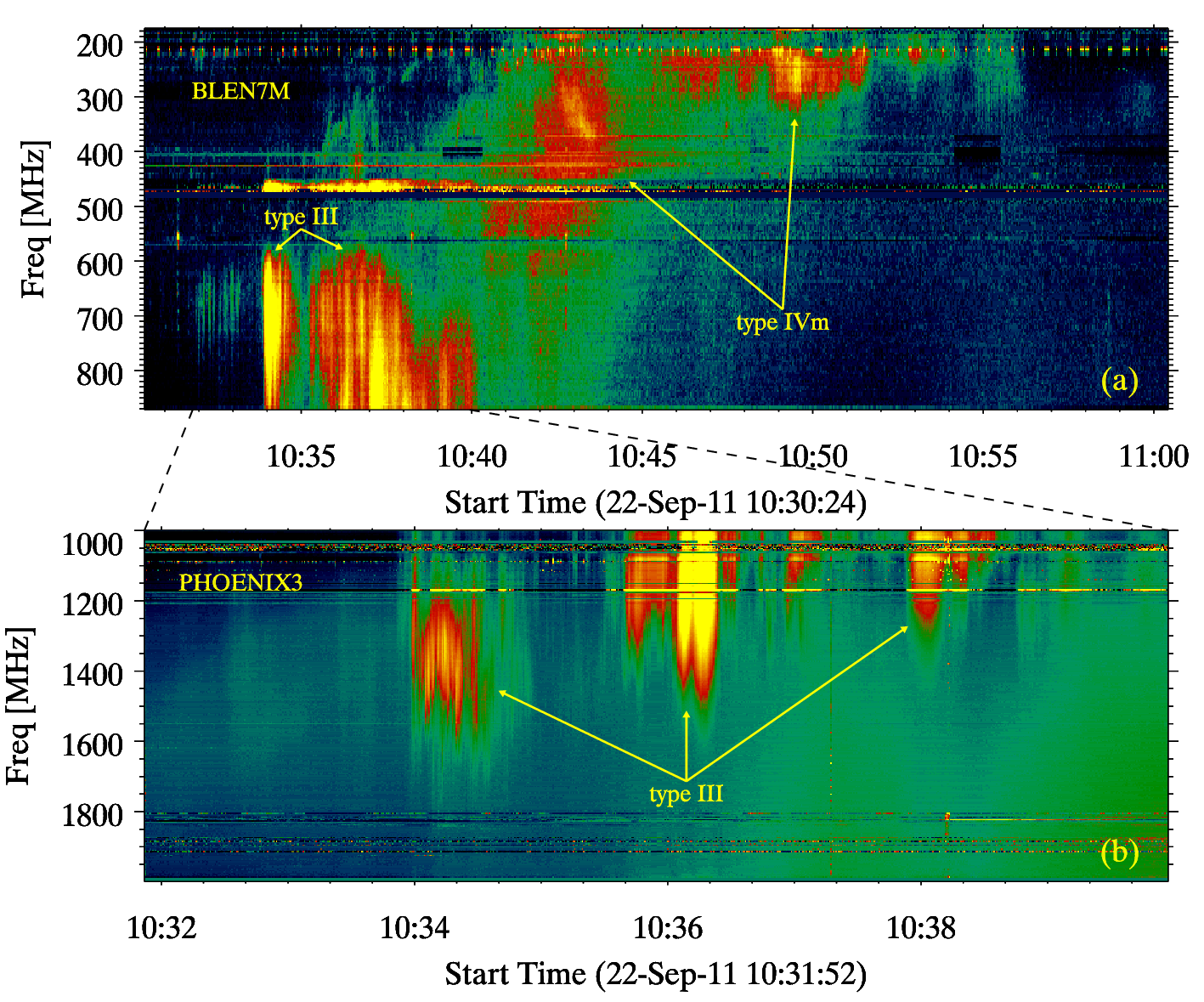}
\centering
\caption{Radio dynamic spectra recorded by e-Callisto/BLEN7M (a) and Ondrejov/PHOENIX3 (b) during the flare impulsive phase.
The yellow arrows point to the type III and IVm bursts.}
\label{fig6}
\end{figure}

Figure~\ref{fig6} shows radio dynamic spectra observed by e-Callisto/BLEN7M during 10:30$-$11:00 UT and Ondrejov/PHOENIX3 during 10:31$-$10:40 UT.
The yellow arrows point to the type III radio bursts during 10:34$-$10:40 UT, with the frequency ranging from $\sim$600 MHz to $\sim$1600 MHz.
The radio fluxes at 720 MHz and 1.40 GHz are extracted from the dynamic spectra and plotted in Figure~\ref{fig1}(e).
The radio spikes during 10:34$-$10:40 UT are roughly coincident with the spikes in GOES derivative flux in 1$-$8 {\AA} as a proxy of HXR flux (Figure~\ref{fig1}(d)).
The occurrence of type III bursts is a signature of intermittent magnetic reconnection and particle acceleration during the flare impulsive phase \citep{kli00,kou22,kar23}.
In Figure~\ref{fig1}(e), the red dotted line delineates the time (10:40:30 UT) when the HC speed gets close to the linear speed of CME, 
implying that acceleration of the HC may come to a halt when flare reconnection slows down.

In Figure~\ref{fig6}(a), apart from the type III bursts, there is a moving type IV (type IVm) burst, whose frequency drifts from $\sim$600 MHz to $\sim$200 MHz during 10:40$-$10:56 UT.
It is generally suggested that a type IVm burst results from emission of electrons trapped in an ascending MFR \citep{bai14,vas19,vem22}.
In our case, as the HC rises up, the electrons are trapped in it and produce type IVm burst.

\section{Energetics} \label{eng}
\begin{figure}
\includegraphics[width=0.45\textwidth,clip=]{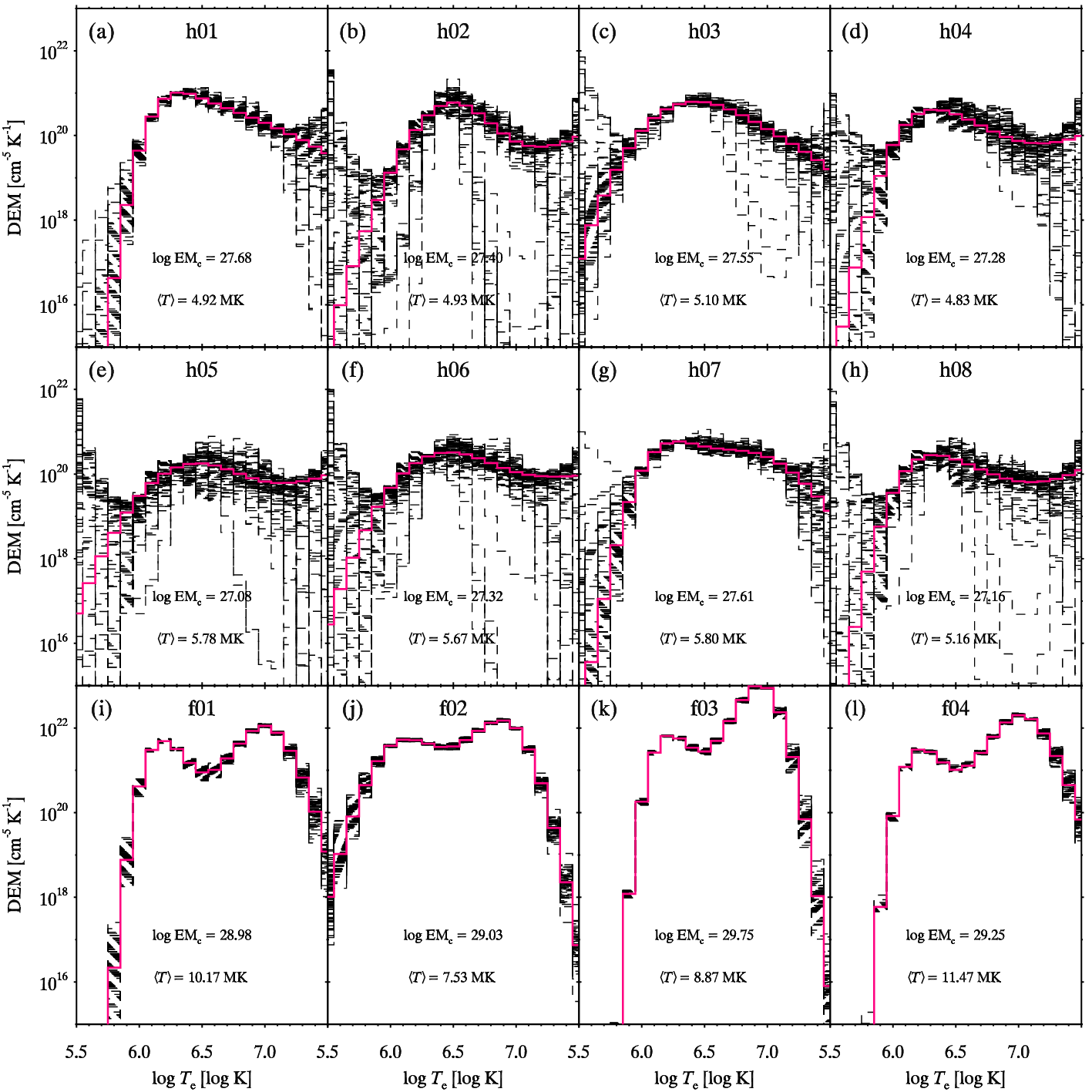}
\centering
\caption{Reconstructed DEM profiles of the eight regions within the HC (a)-(h) and four regions within the flare site (i)-(l).
The red solid lines represent the best-fitted DEM curves from the observed values.
The black dashed lines represent the DEM curves from 100 MC simulations.
The values of $\langle T\rangle$ and $\mathrm{\log EM_{c}}$ are labeled.}
\label{fig7}
\end{figure}

To estimate thermal energy of the HC, we will first calculate its temperature using the DEM analysis \citep{cx12}.
Using simultaneous observations of six AIA passbands (94, 131, 171, 193, 211, and 335 {\AA}),
this method has been widely applied to temperature diagnostics of HCs \citep{cx14}, flare current sheets \citep{xue16}, and coronal jets \citep{zqm14,zqm16}.
The observed intensity of each passband ($I_{j}$) is determined by the temperature response function $R_{j}(T)$ and DEM($T$) of the coronal plasma along the line of sight (LOS):
\begin{equation} \label{eqn-3}
  I_{j}=\int_{T_{1}}^{T_{2}}\mathrm{DEM}(T)\times R_{j}(T)dT,
\end{equation}
where $\log T_{1}$ and $\log T_{2}$ stand for the minimum and maximum temperatures for the integral.
Column emission measure ($\mathrm{EM_{c}}$) along the LOS is:
\begin{equation} \label{eqn-4}
  \mathrm{EM_{c}}=\int_{T_{1}}^{T_{2}}\mathrm{DEM}(T)dT=\int_{0}^{H} n_{e}^{2}dh,
\end{equation}
where $n_{e}$ and $H$ denote the electron number density and LOS depth.
Hence, the DEM-weighted average temperature is:
\begin{equation} \label{eqn-5}
  \langle T\rangle=\frac{\int_{T_{1}}^{T_{2}}\mathrm{DEM}\times T\times dT}{\mathrm{EM_{c}}}.
\end{equation}
Besides, Monte Carlo (MC) simulations are conducted to evaluate the confidence of DEM reconstruction \citep{cx12}.
In Figure~\ref{fig2}(d), we select eight regions (h01$-$h08) within the HC and four regions (f01$-$f04) within the flare site at 10:36:45 UT, which are marked with orange and black boxes.
Reconstructed DEM profiles of the 12 regions are plotted in Figure~\ref{fig7}, with the values of $\langle T\rangle$ and $\mathrm{\log EM_{c}}$ being labeled.
For the HC, there are large uncertainties of DEM profiles at temperatures of $\log T<5.9$ and $\log T>7.1$. 
Therefore, we set $\log T_{1}[\mathrm{K}]=5.9$ and $\log T_{2}[\mathrm{K}]=7.1$ to calculate $\langle T\rangle$ and $\mathrm{EM_{c}}$.
$\langle T\rangle$ lies in the range of 4.8$-$5.8 MK, with a mean value of 5.27 MK and a standard deviation of 0.41 MK. 
Hence, $\langle T\rangle=5.27\pm0.41$ MK (see Table~\ref{tab-1}).
$\mathrm{EM_{c}}$ has a mean value of 2.70$\times10^{27}$ cm$^{-5}$ and a standard deviation of 1.30$\times10^{27}$ cm$^{-5}$, respectively.
For the flare region, the uncertainties of DEM profiles are small at temperatures of $5.5\leqslant \log T \leqslant7.5$.
Therefore, we set $\log T_{1}[\mathrm{K}]=5.5$ and $\log T_{2}[\mathrm{K}]=7.5$.
$\langle T\rangle$ lies in the range of 7.53$-$11.47 MK, with a mean value of 9.51 MK. $\mathrm{EM_{c}}$ is nearly two orders of magnitude larger than that of HC.
Note that the flare temperature recorded by GOES increases and reaches maximum at $\sim$10:44 UT (Figure~\ref{fig1}(b)).

\begin{deluxetable}{cccc}
\tablecaption{Temperature, volume, electron number density, and mass of the HC. \label{tab-1}}
\tablecolumns{4}
\tablenum{1}
\tablewidth{0pt}
\tablehead{
\colhead{$\langle T\rangle$} &
\colhead{$V_{HC}$} &
\colhead{$n_{e}$} &
\colhead{$m_{HC}$} \\
\colhead{(MK)} &
\colhead{(cm$^3$)} &
\colhead{(cm$^{-3}$)} &
\colhead{(g)} 
}
\startdata
5.27$\pm$0.41  & (1.33$\pm$0.03)$\times10^{30}$ & $(5.95\pm1.52)\times10^8$ & $(1.72\pm0.45)\times10^{15}$ \\
\enddata
\end{deluxetable}

The next step is to estimate the volume of the HC. 
We first attempt to fit the HC with the revised graduated cylindrical shell (GCS) model, which is proposed to reconstruct the 3D morphology of prominences or MFRs propagating non-radially \citep{zqm23}. 
Unfortunately, the outcomes are not acceptable after many trials.
Considering that a MFR is a part of a torus \citep{chen89,td99}, we fit the HC with a torus. The parametric equations of a ring torus (or donut) is:
\begin{equation} \label{eqn-6}
\cases{x(\theta,\phi)=(R+r\cos\theta)\cos\phi+x_{0}, \cr
  y(\theta,\phi)=(R+r\sin\theta)\cos\phi+y_{0}, \cr
  z(\theta,\phi)=r\sin\phi+z_{0}, \cr}
\end{equation}
where $0\leq\theta<2\pi$, $0\leq\phi<2\pi$, $R$ and $r$ stand for the radius of the circular axis and radius of the tube ($r<R$), respectively. The volume of such a torus is $V_{rt}=2\pi^2Rr^2$.

In this study, the axis of the HC is not circular during eruption, implying that an ellipse would be more suitable. 
Thereby, the parametric equations of an elliptic torus is:
\begin{equation} \label{eqn-7}
\cases{x(\theta,\phi)=(R_{1}+r\cos\theta)\cos\phi+x_{0}, \cr
           y(\theta,\phi)=(R_{2}+r\sin\theta)\cos\phi+y_{0}, \cr
           z(\theta,\phi)=r\sin\phi+z_{0}, \cr}
\end{equation}
where $R_1$ and $R_2$ represent the semi major axises of the ellipse ($R_{1}<R_{2}$).
The maximal LOS depth of such a tours is $H=2r$ and the total volume is $V_{et}=\pi r^{2}(2\pi R_{1}+4(R_{2}-R_{1}))$.

\begin{figure*}
\includegraphics[width=0.9\textwidth,clip=]{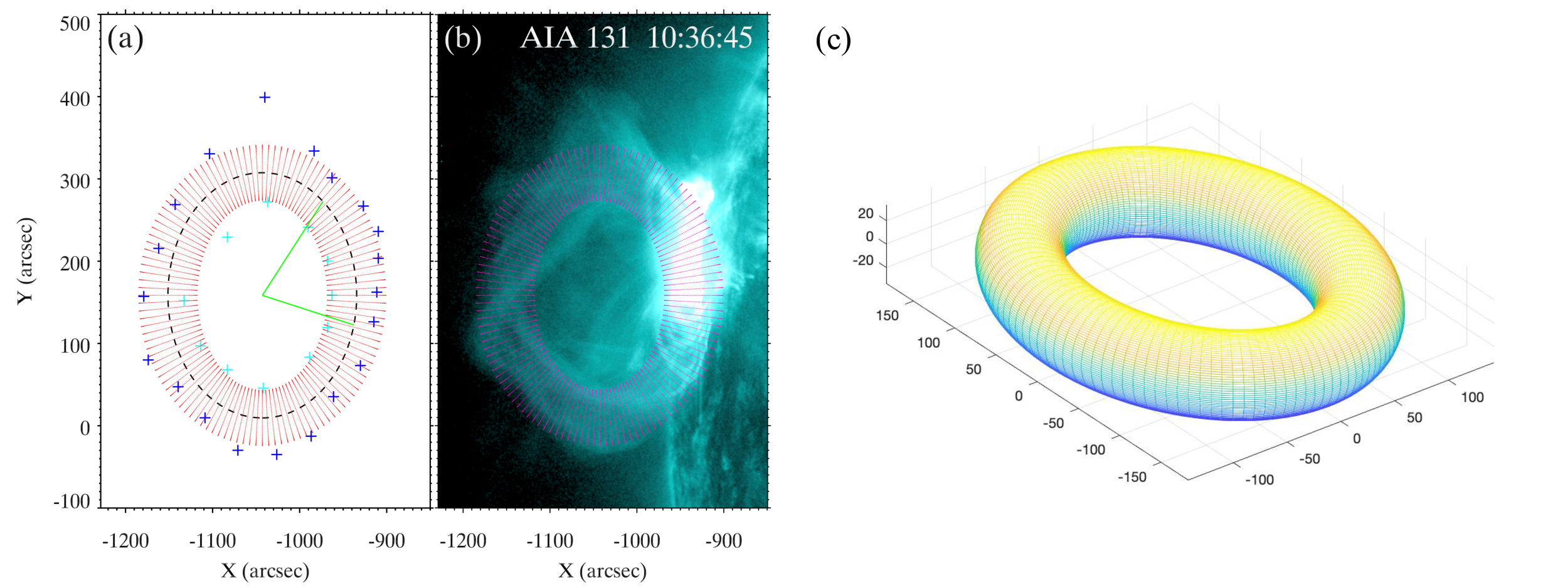}
\centering
\caption{(a) Outer (blue ``+" symbols) and inner (cyan ``+" symbols) boundaries of the HC. The elliptic torus is superposed with red dots. 
The black dashed line denotes the elliptic axis. Two green solid lines are used to separate the flare from HC.
(b) AIA 131 {\AA} image at 10:36:45 UT superposed by the elliptic torus (magenta dots).
(c) 3D visualization of the elliptic torus with $\mathit{Matlab}$.}
\label{fig8}
\end{figure*}

In Figure~\ref{fig2}(d), the outer and inner boundaries of the HC and flare are delineated with magenta and yellow ``+" symbols.
Since the footpoints of the HC may blend with the hot PFLs, we fit the HC and PFLs together with an elliptic torus.
In Figure~\ref{fig8}(a), the same boundaries are delineated with blue and cyan ``+" symbols, respectively.
The fitted elliptic torus is overlaid with red dots in $XY$ plane, 
where $R_{1}=108\farcs4$, $R_{2}=149\farcs0$, $r=33\farcs7$, $x_{0}=-1042\farcs9$, $y_{0}=158\farcs5$, and $z_{0}=0\arcsec$. 
The shorter section between two green solid lines represents the flare region, while the longer section represents the HC.
In Figure~\ref{fig8}(b), the elliptic torus is superposed on the AIA 131 {\AA} image at 10:36:45 UT with magenta dots.
It is obvious that the elliptic torus is in accordance with most of the HC, except for the top part. Figure~\ref{fig8}(c) shows 3D visualization of the whole torus with $\mathit{Matlab}$.
The total volume of the torus is $V_{et}\simeq1.14\times10^{30}$ cm$^3$. 
Considering that the interface between the HC and PFLs is unclear, we pinpoint the interface manually for ten times and calculate the associated volume of HC separately.
It is found that the HC occupies 0.80$\pm$0.02 of $V_{et}$, i.e., (0.91$\pm$0.02)$\times10^{30}$ cm$^3$.

\begin{deluxetable}{c|c}
\tablecaption{Different energy components of the HC, flare, and CME on 2011 September 22. \label{tab-2}}
\tablecolumns{2}
\tablenum{2}
\tablewidth{0pt}
\tablehead{
\colhead{Energy} &
\colhead{Value} \\
\colhead{} &
\colhead{(erg)}
}
\startdata
$E_{tHC}$ & (1.77$\pm$0.61)$\times10^{30}$ \\
$E_{kHC}$ & (2.90$\pm$0.79)$\times10^{30}$ \\
$E_{HC}$ & (4.66$\pm$1.39)$\times10^{30}$ \\
$E_{tFL}$ & (1.63$\pm$0.04)$\times10^{31}$ \\
$E_{rFL}$ & (1.03-1.31)$\times10^{31}$ \\
$E_{nth}$ & (5.44$\pm$0.13)$\times10^{31}$ \\
$E_{kCME}$ & (3.43$\pm$0.94)$\times10^{31}$ \\
$E_{pCME}$ & (2.66$\pm$0.49)$\times10^{30}$ \\
$E_{CME}$ & (3.69$\pm$0.98)$\times10^{31}$ \\
\enddata
\end{deluxetable}

\begin{figure*}
\includegraphics[width=0.90\textwidth,clip=]{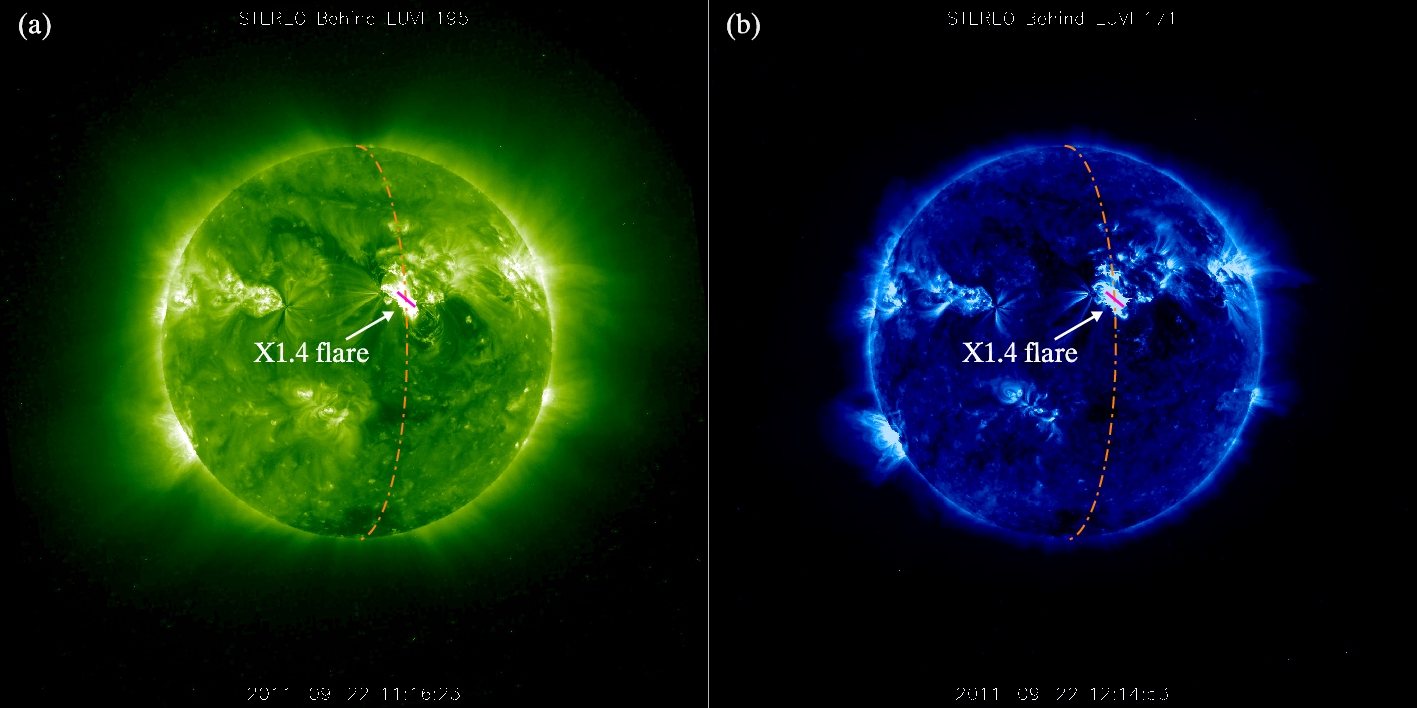}
\centering
\caption{Full-disk STB/EUVI 195 {\AA} image at 11:16:23 UT (a) and 171 {\AA} image at 12:14:53 UT (b).
The orange dot-dashed lines represent the longitude line passing through AR 11302. The short magenta lines represent the PIL.
The arrows point to PFLs of the X1.4 flare.}
\label{fig9}
\end{figure*}

Since AR 11302 is very close to the limb, the tilt angle ($\gamma$) of the polarity inversion line (PIL) in the photosphere with respect to the longitude line is unknown.
As mentioned above, the flare and CME are also captured by the Extreme-UltraViolet Imager \citep[EUVI;][]{how08} on board STB.
Figure~\ref{fig9} shows the full-disk 195 {\AA} image at 11:16:23 UT and 171 {\AA} image at 12:14:53 UT.
The orange dashed lines represent the longitude line passing through AR 11302. The short magenta lines represent the PIL, which is parallel to the flare ribbons.
Assuming that the HC is along the PIL before eruption, the plane of HC deviates from the plane of sky (POS) by $\gamma\simeq47\degr$.
After correcting the projection effect, the true volume of the HC ($V_{HC}$) increases by a factor of $f=(\cos\gamma)^{-1}\simeq1.466$ to (1.33$\pm$0.03)$\times10^{30}$ cm$^3$.
Meanwhile, the LOS depth increases by a factor of $f$, resulting in $H=2rf\simeq71.6$ Mm.
The average electron number density within the HC is $n_{e}=\sqrt{\mathrm{EM_{c}}/H}=(5.95\pm1.52)\times10^8$ cm$^{-3}$ when $\mathrm{EM_{c}}\in[1.4, 4.0]\times10^{27}$.
Assuming a filling factor of unity \citep{kon17}, thermal energy of the HC is:
\begin{equation} \label{eqn-8}
  E_{tHC}=3n_{e}k_{B}\langle T\rangle V_{HC},
\end{equation}
where $k_{B}=1.38\times10^{-16}$ erg K$^{-1}$ is the Boltzmann constant.
Using the above values ($n_e$, $\langle T\rangle$, and $V_{HC}$) including their uncertainties, three arrays with the same number of elements are created separately.
The mean values and standard derivations of these arrays are consistent with the mean values and uncertainties of the physical parameters.
The product of these arrays is employed to obtain the average and standard derivation of $E_{tHC}$, yielding $E_{tHC}=(1.77\pm0.61)\times10^{30}$ erg (Table~\ref{tab-2}).

Assuming the coronal abundance, the total mass of the HC is $m_{HC}=1.3m_{p}n_{e}V_{HC}\simeq(1.72\pm0.47)\times10^{15}$ g, where $m_p$ is the mass of proton.
In Figure~\ref{fig4}(b), $v_{HC}$ increases to $\sim$581 km s$^{-1}$ at 10:36:45 UT. 
The kinetic energy of the HC is $E_{kHC}=\frac{1}{2}m_{HC}v_{HC}^2\simeq(2.90\pm0.79)\times10^{30}$ erg. 
The total energy of the HC, including the thermal and kinetic components, reaches $(4.66\pm1.39)\times10^{30}$ erg, which is close to that of the HC on 2010 November 3 \citep{hk13}.
In Figure~\ref{fig5}, as mentioned above, the CME most probably originates from the eruptive HC. 
Therefore, $m_{HC}$ is considered as a lower limit of the mass of CME ($m_{CME}$) owing to the mass pileup during its propagation.
The velocity of CME ($v_{CME}\simeq1905$ km s$^{-1}$) is derived from linear fitting of the CME height with time.
A lower limit of the kinetic energy of the CME is $E_{kCME}=\frac{1}{2}m_{CME}v_{CME}^2\simeq(3.12\pm0.85)\times10^{31}$ erg.
Assuming an increment of the CME mass by $\sim$10\% during its propagation, $m_{CME}$ reaches $(1.89\pm0.52)\times10^{15}$ g,
which is close to the mass of CME front on 2011 September 13 derived by using the WL observation \citep{ying23}.
In consequence, the kinetic energy of CME increases to (3.43$\pm$0.94)$\times10^{31}$ erg, which is one order of magnitude larger than that of HC (Table~\ref{tab-2}). 

The potential energy of the CME is \citep{li23}: 
\begin{equation} \label{eqn-9}
  E_{pCME}=GM_{\sun}m_{CME}(\frac{1}{R_{\sun}}-\frac{1}{R_{CME}}),
\end{equation}
where $G$ is the gravitational constant, $M_{\sun}$ and $R_{\sun}$ represent the mass and radius of the Sun, $R_{CME}$ denotes the mass center of the CME, which increases with time.
$E_{pCME}$ falls in the range of (2.66$\pm$0.49)$\times10^{30}$ erg during its propagation in the FOV of LASCO, which accounts for $\sim$8\% of the kinetic energy.
The total CME energy, consisting of the kinetic and potential components, is estimated to be (3.69$\pm$0.98)$\times10^{31}$ erg.

Thermal energy of the flare is \citep{kon17,zqm19}:
\begin{equation} \label{eqn-10}
  E_{tFL}=3k_{B}T_{FL}\sqrt{\mathrm{EM_{V}}\times V_{FL}}=3k_{B}T_{FL}\sqrt{\mathrm{EM_{V}}}A_{FL}^{3/4},
\end{equation}
where $T_{FL}$, $A_{FL}$, $V_{FL}=A_{FL}^{3/2}$, and $\mathrm{EM_{V}}$ stand for the temperature, area, volume, and volume emission measure of the flare, respectively.
Figure~\ref{fig1}(b) shows the time evolutions of $T_{FL}$ (orange line) and $\mathrm{EM_{V}}$ (green line) recorded by the GOES spacecraft during 10:20$-$12:20 UT.
The two parameters sequentially reach their maxima at 10:44:13 UT and 11:03:15 UT, while the composite parameter $T_{FL}\sqrt{\mathrm{EM_{V}}}$ reaches maximum at 10:57:08 UT.
Figure~\ref{fig2}(h) shows SXR image of the flare at 6$-$12 keV during 10:54$-$10:58 UT, 
where $A_{FL}$ is defined as the total area of pixels whose intensities are above (0.50$\pm$0.01)$I_M$. Here, $I_M$ denotes the maximum intensity of the SXR image.
After correcting the projection effect by a factor of $f\simeq1.466$, $A_{FL}\simeq(1.69\pm0.06)\times10^{19}$ cm$^2$ and $V_{FL}\simeq(6.95\pm0.37)\times10^{28}$ cm$^3$. 
Therefore, the peak thermal energy of the flare is (1.63$\pm$0.04)$\times10^{31}$ erg.

Figure~\ref{fig1}(c) shows variation of the radiative energy loss rate ($L_{r}(t)$) from SXR-emitting plasma, where background level ($L_b$) is plotted with an olive dashed line.
Hence, the total background-subtracted radiative energy loss is:
\begin{equation} \label{eqn-11}
  E_{rFL}=\int (L_{r}(t)-L_{b}) {dt}.
\end{equation}
The integral during 10:20$-$12:20 UT results in a value of $\sim$1.31$\times10^{31}$ erg.
An alternative way of estimating $E_{rFL}$ based on $\mathrm{EM_V}$ and $\Lambda(T_{FL})$ is \citep{zqm19}:
\begin{equation} \label{eqn-12}
  E_{rFL}=\int\mathrm{EM_V}(t)\times\Lambda(T_{FL}(t))dt,
\end{equation}
where $\Lambda(T_{FL})$ denotes radiative loss rate of the PFLs \citep{ct69,del21}. 
The integral during the same period of time gives rise to a value of $\sim$1.03$\times10^{31}$ erg. Therefore, $E_{rFL}$ is between 1.03$\times10^{31}$ and 1.31$\times10^{31}$ erg.
All the energy components are listed in Table~\ref{tab-2}.

\section{Discussion} \label{dis}
There is a long debate on when a HC is formed. 
Observational evidences are abundant in favor of formation of a HC before eruption \citep{cx13a,zqm22b} as well as during eruption \citep{cx11,song14,gou19,yan21}.
In our study, the HC is obviously formed before eruption, since the onset time of its fast rise is earlier than the start time of flare in Figure~\ref{fig4} \citep[see also][]{cx20,nin20}.
The HC is visible in 131 {\AA} until $\sim$10:40 UT before escaping from the FOV of AIA, implying that a lower limit of the HC lifetime in 131 {\AA} is $\sim$20 minutes. 
Considering the high temperature of HCs, the cooling time due to heat conduction is short \citep{car94}:
\begin{equation} \label{eqn-13}
  \tau_{HC}=4\times10^{-10}\frac{n_{e}L_{HC}^2}{\langle T\rangle^{5/2}},
\end{equation}
where $n_{e}$, $\langle T\rangle$, and $L_{HC}$ represent the average density, temperature, and half length of a HC, respectively.
Using the observed values of $n_{e}=(5.95\pm1.52)\times10^8$ cm$^{-3}$, $\langle T\rangle\simeq5.27\pm0.41$ MK, 
$L_{HC}\simeq(2.50\pm0.06)\times10^{10}$ cm, $\tau_{HC}$ is estimated to be 340$-$4260 s.
The cooling time by radiation is estimated to be $\sim$3.0$\times$10$^5$ s, which is much longer than $\tau_{HC}$.
The short conductive cooling time of a HC suggests that continuous heating is needed to balance the rapid conductive cooling, presumably by magnetic reconnection \citep{hk13}.
\citet{chen19} explored the formation and eruption of a HC on 2014 March 20. It is revealed that heating and buildup of the HC is most probably due to
tether-cutting reconnection at the hyperbolic flux tube in the precursor phase.
Recently, \citet{cx23} investigated the thermal and kinetic evolutions of a hot MFR, especially the slow-rise precursor phase on 2012 March 13. 
It is found that moderate magnetic reconnection at the X-shaped high-temperature plasma sheet is important for the formation and heating of pre-eruptive MFR up to $\sim$8 MK.

During an eruptive flare associated with a CME, the magnetic free energy is impulsively released and mainly converted into radiation, thermal, and nonthermal energies of the flare, 
and kinetic, potential energies of the CME \citep{ems12,asch17}. The energy partitions in flares and CMEs are generally in the same order of magnitude \citep{ems04,ems05,feng13,li23}.
In this study, the peak thermal energy, total radiative energy loss of the flare, and kinetic energy of the CME are in the same order of magnitude ($\sim$10$^{31}$ erg).
It should be emphasized that the bulk kinetic energy of turbulent motions at the flare loop top \citep{kon17} as well as the prompt component of solar energetic particles \citep{li23}
are not taken into account because they contribute only negligible fractions.
Owing to the lack of data from RHESSI before 10:54 UT, we are unable to calculate nonthermal energy of the flare-accelerated electrons ($E_{nth}$).
However, statistical investigation reveals that an excellent correlation exists between the peak thermal energy and nonthermal energy.
The ratio between the two components has a median value of 0.3$\pm$0.02 \citep{war16}.
Hence, $E_{nth}$ is in the range of (5.44$\pm$0.13)$\times10^{31}$ erg, which is sufficient to heat and sustain SXR-emitting plasma in the flare \citep{zqm19}.

In Table~\ref{tab-2}, the thermal and kinetic energies of the HC are a few 10$^{30}$ erg, which are one order of magnitude lower than the flare energies.
Nevertheless, the ratio $\frac{E_{tHC}}{E_{tFL}}\simeq0.11\pm0.03$ is significant, indicating that thermal energy of the HC should be taken into account in energy partition.
The magnetic free energy, which is predominantly stored in a MFR before flare, is estimated to be $\sim$10$^{32}$ erg for a normal X-class flare \citep{guo08}.
Assuming the magnetic field strength ($B$) and volume of the HC are $\sim$150 G \citep{guo21} and (1.33$\pm$0.03)$\times10^{29}$ cm$^3$ before eruption, 
the total magnetic energy is estimated to be $E_{mag}=\frac{B^2}{8\pi}V_{HC}\simeq(1.20\pm0.03)\times10^{32}$ erg.
It is obvious that most of the magnetic free energy is released and distributed into the flare and CME, only a few percent is used to heat the HC.
Intermittent magnetic reconnections in the flare current sheet may play a role in sustaining the hot temperature of the HC.
MHD numerical simulations are especially desired to figure out the formation and thermal evolution of HCs \citep{cx23}.
The elliptic torus shape of HCs (Figure~\ref{fig8}) inspired by theoretical MFR models \citep{chen89,td99} 
is highly beneficial to the calculations of volume, mass, thermal energy, and kinetic energy in statistical analysis.
Multi-point observations of HCs in EUV and SXR passbands, especially from the Sun-Earth Lagrange point L5, 
are expected in next-generation solar telescopes \citep{liuy10,go23} to facilitate 3D reconstructions of HCs.

It should be emphasized that the HC is visible in AIA 131 and 94 {\AA} images until $\sim$10:39 UT. The thermal energy of the X1.4 flare reaches its maximum at $\sim$10:57 UT. 
The CME is visible in the FOV of LASCO during 10:48$-$11:12 UT. The total radiative loss of SXR-emitting plasma is an integral during 10:20$-$12:20 UT. 
Accordingly, it is difficult to compare the energy components at the same instant \citep{ems04,ems05,ems12,feng13,asch17,li23}.
Moreover, as the HC rises and expands, the shape is only regular (resembling an elliptic torus) during 10:35:09$-$10:37:33 UT, when elliptic torus fitting is applicable to the volume estimation.
This is an limitation of the present work. So, there is no need to investigate the temporal evolution of the thermal and kinetic energies of the HC.  
In the future, we will find more cases to study the evolutions in a statistical way.

\section{Conclusion} \label{con}
In this paper, we carry out multiwavelength and multiview observations of the eruption of a HC, which generates an X1.4 flare and a fast CME on 2011 September 22.
We focus on the energetics of the HC and compare with energy components of the related flare and CME.
The main results are as follows:
   \begin{enumerate}
      \item The thermal and kinetic energies of the HC are (1.77$\pm$0.61)$\times10^{30}$ erg and (2.90$\pm$0.79)$\times10^{30}$ erg, respectively.
      The peak thermal energy of the flare and total radiative loss of SXR-emitting plasma are (1.63$\pm$0.04)$\times10^{31}$ erg 
      and (1.03$-$1.31)$\times10^{31}$ erg, respectively.
      The ratio between the thermal energies of HC and flare is 0.11$\pm$0.03, suggesting that thermal energy of the HC is not negligible.
      The kinetic and potential energies of the CME are (3.43$\pm$0.94)$\times10^{31}$ erg and (2.66$\pm$0.49)$\times10^{30}$ erg, 
      giving rise to a total energy of (3.69$\pm$0.98)$\times10^{31}$ erg for the CME.
      \item Continuous heating of the HC is required to balance the rapid cooling by heat conduction, which probably originate from
      intermittent magnetic reconnection at the flare current sheet.
      Our investigation will shed light on the accumulation, release, and conversion of energies in large-scale solar eruptions.
   \end{enumerate}

\begin{acknowledgments}
The authors appreciate the referee for valuable suggestions to improve the quality of this article.
We also thank Drs. Yuandeng Shen, Lei Ni, Zhike Xue, and Xiaoli Yan for inspiring discussions.
SDO is a mission of NASA\rq{}s Living With a Star Program. AIA data are courtesy of the NASA/SDO science teams.
STEREO/SECCHI data are provided by a consortium of US, UK, Germany, Belgium, and France.
The e-Callisto data are courtesy of the Institute for Data Science FHNW Brugg/Windisch, Switzerland.
This work is supported by the Strategic Priority Research Program of the Chinese Academy of Sciences, Grant No. XDB0560000,
the National Key R\&D Program of China 2022YFF0503003 (2022YFF0503000), NSFC under the grant numbers 12373065, 12325303, 12073081, 
and Yunnan Key Laboratory of Solar Physics and Space Science under the grant number YNSPCC202206.
J.D. is supported by the Special Research Assistant Project CAS.
\end{acknowledgments}

\end{document}